\documentclass[showkeys,showpacs,superscriptaddress]{revtex4-2}
\usepackage{amsmath}
\usepackage{amssymb}
\usepackage{xcolor}

\begin{document}

\title{Condensation of a spinor field at the event horizon
}
\author{
Vladimir Dzhunushaliev
}
\email{v.dzhunushaliev@gmail.com}
\affiliation{
Department of Theoretical and Nuclear Physics,  Al-Farabi Kazakh National University, Almaty 050040, Kazakhstan
}
\affiliation{
Institute of Experimental and Theoretical Physics,  Al-Farabi Kazakh National University, Almaty 050040, Kazakhstan
}
\affiliation{
Academician J.~Jeenbaev Institute of Physics of the NAS of the Kyrgyz Republic, 265 a, Chui Street, Bishkek 720071, Kyrgyzstan
}

\author{Vladimir Folomeev}
\email{vfolomeev@mail.ru}
\affiliation{
Institute of Experimental and Theoretical Physics,  Al-Farabi Kazakh National University, Almaty 050040, Kazakhstan
}
\affiliation{
Academician J.~Jeenbaev Institute of Physics of the NAS of the Kyrgyz Republic,
265 a, Chui Street, Bishkek 720071, Kyrgyzstan
}

\begin{abstract}
The physical effect of condensation of a classical spinor field at the event horizon is under consideration. The corresponding solution is sought for the set of the Einstein-Dirac equations. It is shown that in this case  there arises a black hole with a $\delta$-like classical spinor field concentrated at the event horizon.
\end{abstract}


\date{\today}

\maketitle 

\section{Introduction}

In general relativity, there are many solutions describing black holes with matter fields localized on them.
Such solutions do exist because of the possibility to circumvent the no-hair theorem
(see, e.g., Refs.~\cite{Volkov:1998cc,Herdeiro:2015waa,Volkov:2016ehx} and references therein)
and they can be exemplified by the well-known (3+1)-dimensional asymptotically flat static hairy black holes with spherically symmetric
event horizon in the SU(2) Einstein-Yang-Mills theory \cite{Volkov:1989fi,Volkov:1990sva,Bizon:1990sr},
black holes with Skyrmion hairs \cite{Luckock:1986tr,Droz:1991cx,Bizon:1992gb}, and black holes inside magnetic monopoles
\cite{Lee:1991vy,Breitenlohner:1991aa,Breitenlohner:1994di}.

A distinctive feature of the aforementioned solutions is that all such fields localized on black holes have integral spin. In turn, 
to the best of our knowledge, at present there are no solutions describing black holes with fermionic hairs. Although  regular localized solutions of the Einstein-Dirac and
Einstein-Maxwell-Dirac equations are known~\cite{Finster:1998ws,Finster:1998ux,Herdeiro:2019mbz,Herdeiro:2017fhv,Dzhunushaliev:2018jhj,Dzhunushaliev:2019kiy,Herdeiro:2021jgc}, 
it is still impossible to extend these solutions to the case of
finite event horizon: the spinor modes, being gravitationally bound in the black hole spacetime,
decay due to the absence of superradiance mechanism for the Dirac field~\cite{Dolan:2015eua}.
In the present Letter we study a gravitating classical spinor field in general relativity, that is, we seek a solution to a self-consistent set of the Einstein-Dirac equations 
describing a configuration with an event horizon.

\section{Lagrangian and general equations}

The total action for the system under consideration can be written in the form [we use the metric signature $(+,-,-,-)$ and natural units $c=\hbar=1$]
\begin{equation}
\label{action_gen}
	S_{\text{tot}} = - \frac{1}{16\pi G}\int d^4 x \sqrt{-g} R + S_{\text{sp}} .
\end{equation}
Here $G$ is the Newtonian gravitational constant and $R$ is the scalar curvature. 
The action  $S_{\text{sp}}$ for the spinor field $\psi$ appearing in Eq.~\eqref{action_gen} can be obtained from the Lagrangian 
$$
	L_{\text{sp}} =	\frac{\imath}{2} \left(
			\bar \psi \gamma^\mu \psi_{; \mu} -
			\bar \psi_{; \mu} \gamma^\mu \psi
		\right) - \mu \bar \psi \psi ,
$$
where $\mu$ is the mass of the spinor field and the semicolon denotes the covariant derivative defined as
$
\psi_{; \mu} =  [\partial_{ \mu} +1/8\, \omega_{a b \mu}\left( \gamma^a  \gamma^b- \gamma^b  \gamma^a\right)]\psi$. 
Here $\gamma^a$ are the Dirac matrices in the standard representation in flat space  [see, e.g.,  Ref.~\cite{Lawrie2002}, Eq.~(7.27)]. 
In turn, the Dirac matrices in curved space, $\gamma^\mu = e_a^{\phantom{a} \mu} \gamma^a$, are derived  using the tetrad $ e_a^{\phantom{a} \mu}$, and $\omega_{a b \mu}$ is the spin connection
[for its definition, see Ref.~\cite{Lawrie2002}, Eq.~(7.135)].

Then, by varying the action \eqref{action_gen} with respect to the metric and the spinor field, we obtain the Einstein and Dirac equations in curved spacetime:
\begin{eqnarray}
	R_{\mu}^\nu - \frac{1}{2} \delta_{\mu }^\nu R &=&
	8 \pi G \,T_{\mu }^\nu,
\label{feqs_10} \\
	\imath \gamma^\mu \psi_{;\mu} - \mu  \psi &=& 0 ,
\label{feqs_20}\\
	\imath \bar\psi_{;\mu} \gamma^\mu + \mu  \bar\psi &=& 0 .
\label{feqs_30}
\end{eqnarray}
The equation~\eqref{feqs_10} involves the energy-momentum tensor $T_{\mu}^\nu$, which can be written in a symmetric form as
$$
	T_{\mu}^\nu = \frac{\imath }{4}g^{\nu\rho}
	\left[
		\bar\psi \gamma_{\mu} \psi_{;\rho} 
		+ \bar\psi\gamma_\rho\psi_{;\mu} - \bar\psi_{;\mu}\gamma_{\rho }\psi 
		- \bar\psi_{;\rho}\gamma_\mu\psi
	\right] - \delta_\mu^\nu L_{\text{sp}} .
$$
Taking into account the Dirac equations \eqref{feqs_20} and \eqref{feqs_30},
we get, as usually, that the Lagrangian $L_{\text{sp}} = 0 $.

\section{Solution}

Since we consider here only spherically symmetric configurations, it is convenient to choose the spacetime metric in the form
\begin{equation}
	ds^2 = N(r) \sigma^2(r) dt^2 - \frac{dr^2}{N(r)} - r^2 \left(
		d \theta^2 + \sin^2 \theta d \varphi^2
	\right),
\label{metric}
\end{equation}
where $N(r)=1-2 G m(r)/r$, and the function $m(r)$ corresponds to the current mass of the configuration enclosed by a sphere with circumferential radius $r$.

For a description of the spinor field, we take the following stationary {\it Ansatz} 
compatible with the spherically symmetric line element \eqref{metric} (see, e.g., Refs.~\cite{Soler:1970xp,Li:1982gf,Li:1985gf,Herdeiro:2017fhv}):
\begin{equation}
	\psi = 
	\frac{1}{2} e^{-\imath t \Omega }
	\begin{pmatrix}
	 -\imath e^{\frac{1}{2} \imath (\theta -\varphi )} u(r) 	& -e^{\frac{1}{2} \imath (\theta +\varphi )} u(r) \\
	 \imath e^{-\frac{1}{2} \imath (\theta +\varphi )} u(r)  & -e^{-\frac{1}{2} \imath (\theta -\varphi )} u(r) \\
	 -e^{-\frac{1}{2} \imath (\theta +\varphi )} v(r) 	& -\imath e^{-\frac{1}{2} \imath (\theta -\varphi )} v(r) \\
	 e^{\frac{1}{2} \imath (\theta -\varphi )} v(r) 			& -\imath e^{\frac{1}{2} \imath (\theta +\varphi )} v(r) \\
	\end{pmatrix} ,
\label{spinor}
\end{equation}
where $\Omega$ is the spinor frequency and $u(r)$ and $v(r)$ are two real functions. This  {\it Ansatz} ensures that the spacetime remains static. 
Each row of this  {\it Ansatz} describes a  spin-$\frac{1}{2}$ fermion, and these two fermions possess the same masses~$\mu$ and opposite spins. 
Although the energy-momentum tensors of these fermions are not spherically symmetric, their sum gives a spherically symmetric energy-momentum tensor.

Then, substituting the {\it Ansatz} \eqref{spinor} and the metric  \eqref{metric} in the field equations \eqref{feqs_10} and \eqref{feqs_20}, one can obtain the following set of equations:
\begin{align}
	\bar v^\prime + \left(
		\frac{N^\prime}{4 N} + \frac{\sigma^\prime}{2\sigma} 
		+ \frac{1}{\bar{x}}	+ \frac{1}{\bar{x} \sqrt{N}}
	\right) \bar v 
+ \left(
		\frac{1}{\sqrt{N}} - \frac{\bar \Omega}{\sigma N}
	\right) \bar u = & 0,
\label{fieldeqs_1_dmls}\\
	\bar u^\prime + \left(
		\frac{N^\prime}{4 N} + \frac{\sigma^\prime}{2\sigma} 
		+ \frac{1}{\bar{x}}	- \frac{1}{\bar{x}\sqrt{N}}
	\right) \bar u 
	+ \left(
		\frac{1}{\sqrt{N}} + \frac{\bar \Omega}{\sigma N}
	\right) \bar v = & 0,
\label{fieldeqs_2_dmls}\\
	\bar m^\prime = & \bar{x}^2 
	\left[ 
		\frac{\bar \Omega}{\sigma\sqrt{N}}\left(\bar u^2 + \bar v^2\right) 
	\right] ,
\label{fieldeqs_3_dmls}\\
	\frac{\sigma^\prime}{\sigma}	= & \frac{\bar{x}}{N}
	\left[ 
		\frac{2 \bar \Omega}{\sigma\sqrt{N}}\left(\bar u^2 + \bar v^2\right)
		- \left(
			\frac{2 \bar{v} \bar{u}}{\bar{x}} + \bar{u}^2 - \bar{v}^2 
		\right)
	\right] ,
\label{fieldeqs_4_dmls}
\end{align}
where the prime denotes differentiation with respect to the radial coordinate. Here,  Eqs.~\eqref{fieldeqs_3_dmls} and \eqref{fieldeqs_4_dmls} are the  $\left(^t_t\right)$ and  $\left[\left(^t_t\right)~-~\left(^r_r\right)\right]$ 
components of the Einstein equations~\eqref{feqs_10}, respectively. The above equations are written in terms of the following dimensionless variables and parameters:
$$
	\bar{x} =  \mu r, \quad
	\bar \Omega = \frac{\Omega}{\mu}, \quad
	\left(\bar u, \bar v\right) = \sqrt{\frac{4\pi G}{\mu}}\left( u, v\right), \quad \bar{m}=G \mu \,m .  
$$

As we known from the literature, all attempts to find black hole solutions with spinor fields, regular at and outside the horizon, failed. Therefore, we seek another type of solutions: the spinor field is completely concentrated (condensed) at the event horizon at $\bar{x}=\bar{x}_H$, 
that is, this field is described by the $\delta$-like function. 

To begin with, it is necessary to define the notion of the Dirac delta function  in space in the presence of an event horizon.
The reason is that, when integrating over space, a three-volume has the form $dV = \sqrt{-\gamma} d^3x$, where
 $\gamma$ is the determinant of a spatial metric. For the metric \eqref{metric} with the event horizon   $N\left( r_H\right) = 0$ 
 the integration over spatial coordinate $r$ results in the fact that the integral of the Dirac delta function diverges, 
$
		\int_{0}^{\infty} \delta (x)/\sqrt{N(x)}\, dx = \infty ,
$
where $x = \bar{x} - \bar{x}_H$, and so it is necessary to redefine the Dirac delta function so that the integral would be finite.
This can be done as follows: 
\begin{equation}
	\delta_{H}(x) = \delta(x) {\sqrt{N(x)}} . 
\label{delta_Dirac_mod}
\end{equation}

Correspondingly, we seek a solution in the form
\begin{align}
	\bar{u} = & \alpha \delta_{H}\left( x \right) N^{1/4}(x) , 
\label{soln_u}\\
	\bar{v} = &  \beta \delta_{H} \left( x \right) N^{1/4}(x) , 
\label{soln_v}
\end{align}
where $\alpha$ and $\beta$ are some constant parameters. For any other choice of the exponent $\gamma$ of the factor
$N^{\gamma}$ in the expressions \eqref{soln_u} and \eqref{soln_v} we will have either zero fermionic charge
 $Q = 0$ for $\gamma > 1/4$ or infinite  $Q = \infty$ for $\gamma < 1/4$. Thus, the value  $\gamma = 1/4$ is the only one for which the fermionic charge
 $Q$ is a finite number.  
Since Eqs.~\eqref{fieldeqs_1_dmls}-\eqref{fieldeqs_4_dmls} contain nonlinear terms of the type of $\bar{v}^2, \cdots$, this results in the appearance of squares  and cubes  $\delta^2(x), \delta^3(x)$ 
in the corresponding equations. Correspondingly, it is necessary to check the mathematical correctness of the expressions containing such quantities. To do this, we choose the following  $\epsilon$-approximation form for the Dirac delta function (with $\epsilon\to 0$):
\begin{equation*}
	\left( \delta_{H}\right)_\epsilon (x) 
	=  \frac{1}{\sqrt{\pi} \epsilon} e^{-\frac{ x^2}{\epsilon^2}} \sqrt{N(x)}, 
\end{equation*}
and assume that in the vicinity of the horizon  (as $x\to 0$) the function $N(x)$ behaves as
\begin{equation*}
		N(x) =  N_H  x + \cdots , 
\end{equation*}
where $N_H$ is some constant parameter; below we will show that this choice for the function $N(x)$ is consistent with the solution of the Einstein equations.

Apparently, there are no solutions with $\bar\Omega \neq 0$; therefore, in Eqs.~\eqref{fieldeqs_1_dmls}-\eqref{fieldeqs_4_dmls}, we henceforth set $\bar\Omega = 0$. As a result, the Dirac equations~\eqref{fieldeqs_1_dmls} and \eqref{fieldeqs_2_dmls} take the following $\epsilon$-form: 
\begin{align}
	\left( \delta_{H}\right)_\epsilon (x)
	\left[ 
		\sqrt{N_H} \left( \alpha + \frac{\beta}{x_H}\right) x^{1/4} 
		+ \beta \frac{\sqrt{N_H}}{x_H} x^{3/4} 
		+ \left( 1 - 2 \frac{x^2}{\epsilon^2}\right) \frac{N_H \beta}{x^{1/4}} 
	\right] & 
\nonumber \\
	- \left( \delta_{H}\right)_\epsilon \left(\sqrt{3} \, x\right)
	 \frac{N_H^{3/2} }{4 \pi \sqrt{3}} \beta 
	\left( 
		x_H \alpha^2 + 2 \alpha \beta - x_H \beta^2 
	\right) \frac{x^{5/4}}{\epsilon^2} 
	& = 0 , 
\label{Dirac_eps_1}\\
	\left( \delta_{H}\right)_\epsilon (x)
	\left[ 
		\sqrt{N_H} \left( \beta - \frac{\alpha}{x_H}\right) x^{1/4} 
		+ \beta \frac{\sqrt{N_H}}{x_H} x^{3/4} 
		+ \left( 1 - 2 \frac{x^2}{\epsilon^2}\right) \frac{N_H \alpha}{x^{1/4}}
	\right] & 
\nonumber \\
	- \left( \delta_{H}\right)_\epsilon \left(\sqrt{3} \, x\right)
	 \frac{N_H^{3/2} }{4 \pi \sqrt{3}} \alpha  
	\left( 
		x_H \alpha^2 + 2 \alpha \beta - x_H \beta^2 
	\right) \frac{x^{5/4}}{\epsilon^2} 
	& = 0 .
\label{Dirac_eps_2}
\end{align}
Here we used the fact that all the functions have $\delta$-like form; this enabled us to change  $\bar x$ in the terms 
$
\frac{1}{\bar x}	\pm \frac{1}{\bar x \sqrt{N}}
$
in the Dirac equations~\eqref{fieldeqs_1_dmls} and \eqref{fieldeqs_2_dmls} by $\bar x_H$: 
$
\frac{1}{\bar x_H}	\pm \frac{1}{\bar x_H \sqrt{N}}
$. Also,  we substituted here the quantity  
$
	\sigma^\prime/\sigma
$
from the Einstein equation~\eqref{fieldeqs_4_dmls}. 

First of all, let us  recall that all expressions in Eqs.~\eqref{Dirac_eps_1} and \eqref{Dirac_eps_2} 
are distributions with the support located at the event horizon. This means that, 
in order to understand the behaviour of these  distributions, it is necessary to integrate the corresponding expression and to examine 
the behaviour of the resulting quantity in the limit $\epsilon \rightarrow 0$. 
We multiply the Dirac equations \eqref{Dirac_eps_1} and \eqref{Dirac_eps_2} by $x^{1/4}$ and consider the behaviour of the first two terms
in these equations after integration as $\epsilon \rightarrow 0$: the first term is $\sim \epsilon$, the second term is $\sim \epsilon^{3/2}$ 
[for details, see Appendix \ref{appendix}, Eq.~\eqref{app_10}]. This means that the corresponding  distributions vanish
in the limit $\epsilon \rightarrow 0$. The third term in these equations, after integration, does no depend on $\epsilon$ 
and is identically equal to zero  [for details, see Appendix \ref{appendix}, Eqs.~\eqref{app_20} and \eqref{app_30}]. 
We set the last terms in these Dirac equations to zero by requiring
\begin{equation}
	x_H = \frac{2 \alpha \beta}{\beta^2 - \alpha^2} .
\label{zero_out}
\end{equation}
It is interesting to note here that the radius of the event horizon can be made as large as desired for $\beta^2 \rightarrow \alpha^2$.

Thus, we may say that the left-hand side of the Dirac equations \eqref{Dirac_eps_1} and \eqref{Dirac_eps_2}, 
after multiplying by  $x^{1/4}$, is a distribution (the third term in these equations),
\begin{equation}
	\mathcal{D}_{\alpha, \beta} = N_H  \lim\limits_{\epsilon \rightarrow 0} 
	\left[ 
		\left( \delta_{H}\right)_\epsilon  (x) \left( 1 - 2 \frac{x^2}{\epsilon^2}\right) 
	\right] 
	\begin{pmatrix}
		\alpha \\
		\beta
	\end{pmatrix} 
	= N_H  \mathcal{F}(x) 
	\begin{pmatrix}
		\alpha \\
		\beta
	\end{pmatrix} , 
\label{Dirac_distr}
\end{equation}
where the definition of the distribution $\mathcal{F}(x)$ is given in Appendix \ref{appendix}, see Eqs. \eqref{app_20} and \eqref{app_30}. 
This distribution vanishes in the sense that
\begin{equation}
	\int \limits_{0}^\infty \frac{\mathcal{D}_{\alpha, \beta}}{\sqrt{N(x)}} dx = 0 . 
\label{LHS_Dirac_zero}
\end{equation} 
This enables us to conclude that, by choosing the function $\mathcal{D}_{\alpha, \beta}$ in the form \eqref{Dirac_distr}, 
the Dirac equations \eqref{fieldeqs_1_dmls} and \eqref{fieldeqs_2_dmls}  are satisfied upon substituting the functions
 $\bar{u}$ and $\bar{v}$ in the form \eqref{soln_u} and \eqref{soln_v}. Note once again that the vanishing of the equations
 \eqref{Dirac_eps_1} and \eqref{Dirac_eps_2} must be understood in the sense of distributions, that is, after integration over $x$.

To calculate  a fermionic charge $Q$, consider the behaviour of the local charge density near the horizon,
\begin{equation}
	\sqrt{-g} j^t = \sqrt{-\gamma} \psi^\dagger \psi 
	= \frac{1}{4\pi G\mu}\frac{\sin \theta}{\sqrt{N}}\bar{x}^2\left(\bar{u}^2+\bar{v}^2\right)
	= \frac{\alpha^2 + \beta^2}{4\pi^2 G\mu \epsilon^2} 
	 N_H x\left(x+\bar{x}_H\right)^2 e^{-\frac{ 2 x^2}{\epsilon^2}}\sin \theta .
\label{charge_dens}
\end{equation}
Here we take into account that the exponential $e^{-\frac{ 2 x^2}{\epsilon^2}}$ is localised near the event horizon within a region of width 
$\sim \epsilon$ and hence $N(x) \rightarrow N\left(\bar{x}_H\right)$. This observation leads to the conclusion that  the local charge density $\sqrt{-g} j^t$, 
in the limit $\epsilon \rightarrow 0$, is localised at the event horizon. This, in turn, implies that,  from the point of view of the quantum Dirac equation, the 
probability of finding a fermion in space for $r > r_H$ is equal to zero.  Note also that 
the function $\frac{x e^{-\frac{ 2 x^2}{\epsilon^2}} }{\epsilon^2} $ should be treated as a distribution, and its properties are discussed in
Appendix \ref{appendix}, see Eq.~\eqref{app_40}.

Let us now consider a fermionic charge. Taking into account Eq.~\eqref{charge_dens}, we have
\begin{equation*}
\begin{split}
	Q & =	\int \sqrt{-g} j^t d^3 x = 
	\int \sqrt{-\gamma} \sqrt{g_{00}} e^t_{\phantom{t}0} \bar{\psi} \gamma^0 \psi d^3 x = 
	\int \sqrt{-\gamma} \psi^\dagger \psi d^3 x = 
	\frac{1}{G\mu^2} \int \limits_{0}^{\infty} \frac{\bar{u}^2 + \bar{v}^2}{\sqrt{N}} \left(x+\bar{x}_H\right)^2 d x 
\nonumber \\
	& \xrightarrow{\text{near the horizon}}  \frac{\alpha^2 + \beta^2}{\pi G\mu^2} N_H \bar{x}_H^2
	\int \limits_{0}^{\infty} \frac{x}{\epsilon^2} e^{-\frac{ 2 x^2}{\epsilon^2}} dx 
	= \frac{\alpha^2 + \beta^2}{4\pi G\mu^2} N_H \bar{x}_H^2 . 
\end{split}
\end{equation*}
From this we can conclude that the charge density is $\delta$-localised at the event horizon and the fermionic charge is nonzero.

Next, keeping in mind that $\bar{\Omega}=0$, the integration of Eq.~\eqref{fieldeqs_3_dmls} yields
\begin{equation}
	\bar m = \bar m_H=\text{const.} 
\label{m_soln}
\end{equation}
This enables us to determine the metric function $N(x)$ in the explicit form
\begin{equation}
	N(x) = 1 - \frac{2 \bar m_H}{\bar x} = \frac{\bar x - \bar x_H}{\bar x_H} + \cdots ,
\label{N_func}
\end{equation}
where the last expression is a Taylor expansion near the event horizon, 
which justifies the choice of the {\it Ansatz}  \eqref{soln_u} and \eqref{soln_v} for the functions $\bar{u}$ and $\bar{v}$.  This means that
$$
	\bar x_H = 2\bar m_H . 
$$

The right-hand side of Eq.~\eqref{fieldeqs_4_dmls} with the condition \eqref{zero_out} is equal to zero, and hence
the solution of this Einstein equation is
\begin{equation}
	\sigma = \text{const.} = 1. 
\label{sigma_soln}
\end{equation}

\section{Discussion and conclusions}

We have shown that a gravitating classical spinor field can create a black hole whose properties differ in principle from the properties of black holes supported by a $\delta$-like mass (a Schwarzschild black hole) or by fields with integral spin (a Reissner-Nordstr\"om black hole, non-Abelian black holes, etc.). This principle difference consists in the fact that in the present case the spinor field is concentrated at the event horizon, in contrast to Reissner-Nordstr\"om and non-Abelian black holes. In some sense, one can say that in the presence of the event horizon a spinor field is condensed on it. This emphasizes once again the uniqueness of the event horizon in general relativity: apart from the fact that it is a trapping surface, i.e., no signal can escape the region of space  located inside the event horizon, it (the event horizon) condensates a classical spinor field.

Let us enumerate the results obtained: 
\begin{itemize}
\item Self-consistent solution to the Einstein-Dirac equations describing a black hole supported by a gravitating classical spinor field is obtained.
\item Solution of the Dirac equation (zero mode) is obtained in terms of distribution functions  with the support located at the event horizon.
\item It is shown that the left-hand side of the Dirac equations vanishes in the sense of distributions; this allows one to consider the Dirac equation as being satisfied in the sense of distributions.
\item The spinor field, which is a source of the black hole, has a $\delta$-like structure and is concentrated (condensed) at the event horizon. 
This allows one to conclude that the presence of an event horizon is the physical mechanism responsible for the condensation of fermions (a fermionic cloud) on this distinguished surface.
\end{itemize}

As we known from the literature, all attempts to find black hole solutions with spinor fields, regular at and outside the horizon, failed. 
We believe that the reason for this fact is the presence of an event horizon, which gives rise to singular terms of the form
$1/N, 1/\sqrt{N}$ in the Dirac equations \eqref{fieldeqs_1_dmls} and \eqref{fieldeqs_2_dmls}. The terms $1/\sqrt{N}$ cannot cancel each other
simultaneously in the Dirac equations \eqref{fieldeqs_1_dmls} and \eqref{fieldeqs_2_dmls}, since they enter the equations in an asymmetric manner. 
The same is true for the terms $1/N$. The appearance of these terms is due to the presence of the spinor frequency $\Omega$, 
as well as of the spin connection  $\omega_{a b \mu}$. In our case the terms with $\Omega = 0$ vanish because we consider the zero mode, whereas the remaining terms are canceled by the corresponding distributions.

After obtaining this solution, there arises a number of interesting questions requiring further investigation. It is known that there are no classical spinor fields in nature (at least now we do not know any manifestations of such fields). Then the question arises: how the nature of such a black hole will have changed (or not changed) if it will be created by a quantized spinor field? In this connection it is worth mentioning a hypothetical effect of the Hawking radiation at the event horizon of a black hole. It is known that this effect arises due to quantum field fluctuations near the event horizon. Keeping in mind that in the present study we consider a classical spinor field at the event horizon, the question arises: whether or not the effect of condensation of the classical spinor field at the event horizon is related to the effect of the Hawking radiation for a quantum field?

To conclude, let us enumerate some questions appearing after the study performed here:
\begin{itemize}
\item Whether black holes supported by a spinor field with $\Omega \neq 0$ do exist? 
\item How the properties of a black hole with a spinor field will have changed after inclusion of fields with integral spin?
\item What happens if the classical spinor field is replaced by a quantum one? Perhaps it is equivalent to a consideration of the Hawking radiation 
on account of backreaction of such radiation on the black hole; i.e., it is a self-consistent problem of existence of a dynamical black hole creating the Hawking radiation. 
\end{itemize}

\section*{Acknowledgements}

We gratefully acknowledge support provided by the program
No.~AP26195069 (Bound states in Maxwell, Yang-Mills, Proca theories with and without gravity in the presence of spinor fields)
of the Committee of Science of the Ministry of Science and Higher Education of the Republic of Kazakhstan.
We also wish to thank an anonymous referee whose comments helped us improve the manuscript.

\appendix

\section{Mathematical properties of the functions appearing in the Dirac equation}
\label{appendix}

For performing calculations, it is necessary to check the mathematical correctness of the right-hand sides of the equations~\eqref{fieldeqs_3_dmls} and \eqref{fieldeqs_4_dmls}, as well as of all terms in the remaining equations~\eqref{fieldeqs_1_dmls} and  \eqref{fieldeqs_2_dmls}. 

After multiplying Eqs.~\eqref{Dirac_eps_1} and \eqref{Dirac_eps_2} by $x^{1/4}$, there are the distributions 
$
	\lim\limits_{\epsilon \rightarrow 0} 
		\left[\left( \delta_{H}\right)_\epsilon (x) x^{\gamma}
	\right] 
$ with $\gamma = 1/2, 1$. 
By definition, distributions are linear functionals, i.e., they are given by the corresponding integrals over the three-volume.
The corresponding integrals for $\epsilon$-approximations of these functions are
\begin{equation}
	\int \limits_{0}^{\infty} \left( \delta_{H}\right)_\epsilon (x) \frac{x^\gamma}{\sqrt{N_H x}} dx 
	= \frac{\Gamma \left(\frac{1+\gamma}{2}\right)}{2 \sqrt{\pi }} 
	\epsilon^{\gamma } \stackrel{\epsilon \rightarrow 0}{\longrightarrow} 0 .
\label{app_10}
\end{equation}
This means that the distribution
$
	\lim\limits_{\epsilon \rightarrow 0} 
		\left[\left( \delta_{H}\right)_\epsilon (x) x^{\gamma}
	\right] = 0 . 
$

Consider the properties of the distribution
$
	\mathcal{F}(x) = 
	\lim\limits_{\epsilon \rightarrow 0} 
	\left[ 
		\left( \delta_{H}\right)_\epsilon (x) \left( 1 - 2 \frac{x^2}{\epsilon^2}\right) 
	\right] 
$. Analogously to the previous function, its properties in the $\epsilon$-approximation are determined by the integral 
\begin{equation}
	\int \limits_{0}^{\infty} 
	\left( \delta_{H}\right)_\epsilon (x) \left( 1 - 2 \frac{x^2}{\epsilon^2}\right) \frac{dx}{\sqrt{N_H x}} = 0 .
\label{app_20}
\end{equation}
This means that the properties of the distribution $\mathcal{F}$ are determined by the following limiting transition:
\begin{equation}
	\int \limits_{0}^{\infty} \mathcal{F}(x) dx = \lim\limits_{\epsilon \rightarrow 0} 
	\left[ 
		\int \limits_{0}^{\infty} 
		\left( \delta_{H}\right)_\epsilon (x) \left( 1 - 2 \frac{x^2}{\epsilon^2}\right) \frac{dx}{\sqrt{N_H x}}
	\right] = 0 .
\label{app_30}
\end{equation}
That is, it vanishes in the sense of distributions.

 Consider the properties of the distribution 
$
	\mathcal{G}(x) = 
	\lim\limits_{\epsilon \rightarrow 0} 
	\left[ 
		\frac{x e^{-\frac{ 2 x^2}{\epsilon^2}} }{\epsilon^2} 
	\right] 
$. Analogously to the function $\mathcal{F}$, its properties in the $\epsilon$-approximation are determined by the integral 
\begin{equation}
	\int \limits_{0}^{\infty} \frac{x e^{-\frac{ 2 x^2}{\epsilon^2}} }{\epsilon^2} dx 
	= \frac{1}{4  },
\label{app_40}
\end{equation}
and therefore  
\begin{equation}
	\int \limits_{0}^{\infty} \mathcal{G}(x) dx = \lim\limits_{\epsilon \rightarrow 0} 
	\left[ 
		\int \limits_{0}^{\infty} 
		\frac{x e^{-\frac{ 2 x^2}{\epsilon^2}} }{\epsilon^2} dx
	\right] 
	= \frac{1}{4 }.
\label{app_50}
\end{equation}
This means that the distribution  $\mathcal{G}(x)$ is nonzero.

\end{document}